\documentclass[mathleft
]{an}
\usepackage{graphicx}
\usepackage{times}
\begin{document}

\Pagespan{789}{}

\title{The coronal parameters of local Seyfert galaxies}

\author{A. Marinucci\inst{1}\fnmsep\thanks{Corresponding author:
  \email{marinucci@fis.uniroma3.it}\newline}
\and A. Tortosa\inst{1}, on behalf of the NuSTAR AGN Physics Working Group
}
\titlerunning{The coronal parameters of local Seyfert galaxies	}
\authorrunning{A. Marinucci \& A. Tortosa}
\institute{
Dipartimento di Matematica e Fisica, Universit\`a degli Studi Roma Tre, via della Vasca Navale 84, 00146 Roma, Italy
}



\abstract{%
   One of the open problems for AGN is the nature of the primary X-ray  emission: it is likely due to Comptonization of soft UV photons, but the optical depth and temperature of the emitting corona were largely unknown before the launch of the Nuclear Spectroscopic Telescope Array ({\it NuSTAR}). It is the first focusing hard X-ray telescope on orbit, ~100 times more sensitive in the 10-79 keV band compared to previous observatories, enabling the study of AGN at high energies with high precision. We present and discuss the results on the hot corona parameters of Active Galactic Nuclei that have been recently measured with {\it NuSTAR} (often in coordination with XMM-{\it Newton}, {\it Suzaku} or {\it Swift}) with unprecedented accuracy, in a number of local Seyfert galaxies.
  }

\maketitle

\section{Introduction}
Bright Active Galactic Nuclei (AGN) are believed to host a supermassive black hole at their center, surrounded by a geometrically thin accretion disc. One of the main open problems for AGN is the nature of the primary X-ray emission: it is observed as a hard power law continuum that dominates the spectral emission above 2 keV. It is thought to arise in a corona above the accretion disc composed of a plasma of hot electrons, due to inverse Compton of UV/optical seed photons from the disc. The two-phase model (Haardt \& Maraschi 1991, Haardt et al. 1994) postulates that assuming thermal equilibrium, a plane-parallel geometry for the electron plasma and neglecting direct heating of the disc, the average spectral properties of Seyfert galaxies in the X-rays could be naturally accounted for.

This physical process produces a typical power-law spectrum extending to energies determined by the electron temperature in the hot corona (Rybicki \& Lightman 1979). The power-law index is function of the plasma temperature $\rm kT_e$ and optical depth $\tau$, while the cutoff energy is mainly related to the former. However, the geometry, optical depth and temperature of the scattering electrons are largely unknown. Comptonization models imply that E$\rm {}_c\simeq$2--3$\rm \times kT_e$ (Petrucci et al. 2000, 2001) and hence simultaneously  measuring the slope and the cutoff energy of the primary power law is fundamental for determining the properties of the hot corona. In this scenario, Fabian et al. 2015 stressed the importance that pair production and annihilation may play in measuring such coronal parameters.
\begin{figure*}
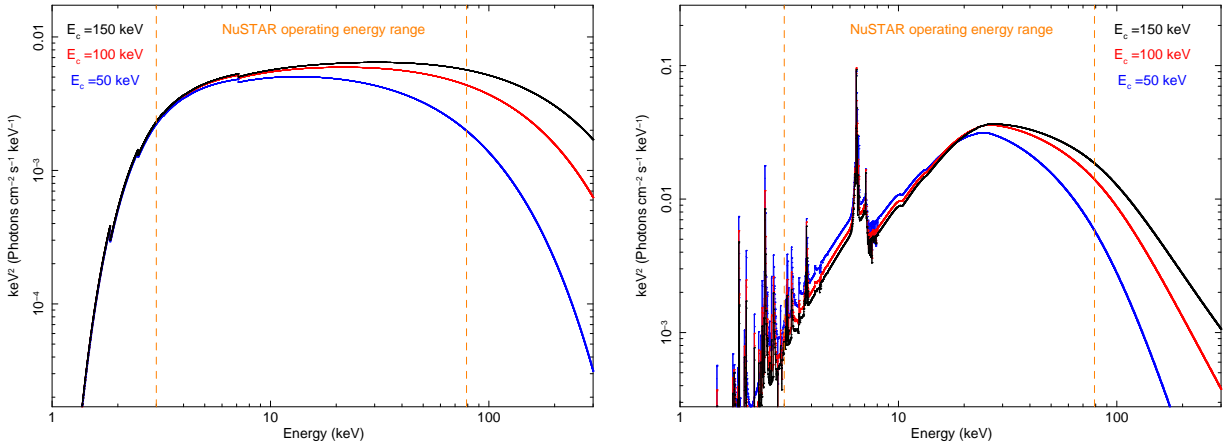

\includegraphics[width=0.27\paperwidth, angle=-90]{pow_new.ps}
\includegraphics[width=0.27\paperwidth, angle=-90]{xill_new.ps}
\caption{{\it Left-hand panel:} Power-law models with $\Gamma=1.8$ absorbed by a column density N$\rm _H=5\times 10^{22}$ cm$^{-2}$ with different cutoff energies (E$\rm _c$=50, 100, 150 keV) for a source with a 2--10 keV flux F$_{2-10}=10^{-11}$ erg cm$^{-2}$ s$^{-1}$. {\it Right-hand panel:} Reflection models with $\Gamma=1.8$, ionization parameter $\xi=100$ erg cm s$^{-1}$ and solar abundances,  with different cutoff energies (E$\rm _c$=50, 100, 150 keV), for a source with a 2--10 keV flux F$_{2-10}=10^{-11}$ erg cm$^{-2}$ s$^{-1}$. Vertical dotted orange lines indicate the energy band in which {\it NuSTAR} operates.} 
\label{models}
\end{figure*}

To investigate the shape of the nuclear continuum, this has to be disentangled from the other complex spectral features in this energy range such as reflection from the accretion disc and from distant material. Typical X-ray features of the cold circumnuclear material include an intense Fe K$\alpha$ line at 6.4 keV due to fluorescence emission and the associated Compton reflection continuum peaking at $\sim$30 keV (Matt et al. 1991; George \& Fabian 1991). These features can be attributed to the reprocessing of the nuclear radiation by distant, neutral matter  (i.e. the 'pc-scale torus': Antonucci 1993). 
Furthermore, there are several important spectral features, like the warm absorber, the soft excess, and the relativistic component of the iron K$\alpha$ line, which are present in a number of objects. 

In the past few years, several cutoff energies in nearby Seyfert galaxies have been measured with hard X-ray satellites, such as {\it BeppoSAX} (Dadina et al. 2007, Perola et al. 2002) and {\it INTEGRAL} (Panessa et al. 2011, De Rosa et al. 2012, Molina et al. 2013). These measurements ranged between 50 and 300 keV and the lack of focusing instruments at high energies resulted in large uncertainties and degeneracy between the cutoff energies and other physical observables (in particular the slope of the primary power law and the amount of radiation Compton scattered by circumnuclear matter).
  
{\it NuSTAR} (Harrison et al., 2013) was launched in June 2012 and has been an observational breakthrough in X-ray astronomy with its unprecedented sensitivity at high energies, operating in the 3-79 keV energy range. It is the first high energy focusing X-ray telescope on orbit, $\sim$100 times more sensitive in the 10--79 keV band compared to previous observatories covering these energies, enabling the study of AGN at high energies with high precision. Simultaneous observations with other X-ray observatories operating below 10 keV, such as XMM-{\it Newton}, {\it Suzaku} and {\it Swift} allowed to measure cutoff energies with great accuracy in a number of sources: MCG-5-23-15 ($116^{+6}_{-5}$ keV: Balokovi\'c et al.2015), SWIFT J2127.4+5654 ($108^{+11}_{-10}$ keV: Marinucci et al.2014a), IC 4329A ($186\pm14$ keV: Brenneman et al. 2014), 3C382 ($214^{+147}_{-63}$ keV: Ballantyne et al. 2014).

In the following, we present and discuss recent results on coronal parameters of Active Galactic Nuclei that have been measured with {\it NuSTAR} in its first three years of science, in a number of local Seyfert galaxies.
\begin{figure*}
\centering
\includegraphics[width=\columnwidth, angle=-90]{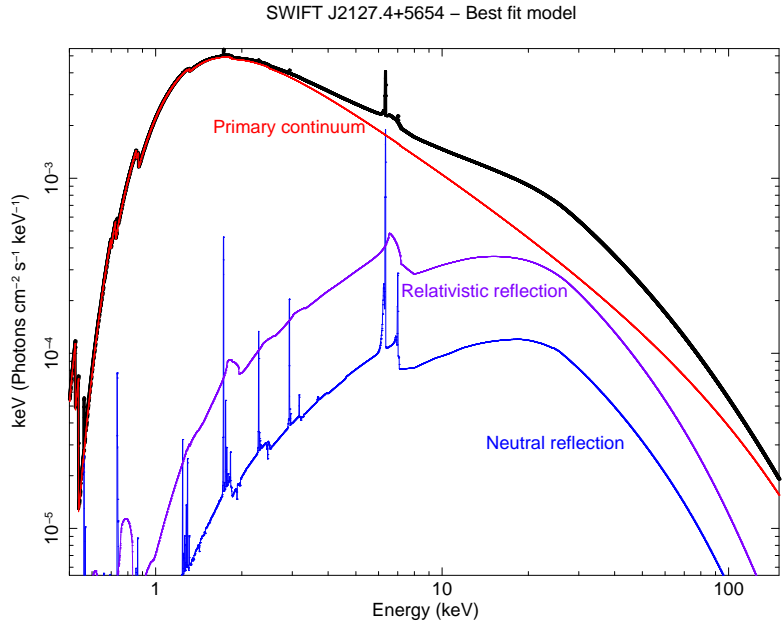}
\caption{Best fit model of the 2013 XMM+{\it NuSTAR} data set of the Narrow Line Seyfert 1 SWIFT J2127.4+5654 (Marinucci et al. 2014a) in which the three main spectral components can be seen: the nuclear continuum (in red), the relativistic reflection from the inner regions of the disc (in purple) and the neutral reflection arising from distant material.}
\label{model}
\end{figure*}

\begin{table*}
 \centering
\label{tlab}
\begin{tabular}{ccccc}
Name & $\Gamma$ & E$\rm _c$ & $\log \rm (M)$ & Reference \\
 & & (keV)&(M$_{\odot}$) & \\
\hline
& & & & \\
3C 382 & $1.68^{+0.03}_{-0.02}$& $214^{+147}_{-63}$& $9.2\pm0.5$& 1-2\\
3C 390.3 & $1.70\pm0.01$& $116^{+24}_{-8}$& $8.4\pm0.1$& 2-12\\
Ark 120 & $1.73\pm0.02$& $>190$& $8.2\pm0.1$& 3-12\\
IC 4329A & $1.73\pm0.01$& $186\pm14$& $6.8\pm0.2$& 4-13\\
Fairall 9 & $1.96^{+0.01}_{-0.02}$& $>242$& $8.4\pm0.1$& 2-12\\
MCG 5-23-16& $1.85\pm0.01$& $166^{+6}_{-5}$& $7.8\pm0.2$& 5-13\\
MCG 6-30-15 &$2.061\pm0.005$& $>110$ & $6.2\pm 0.1$& 6-13\\
Mrk 335& $2.14^{+0.02}_{-0.04}$& $>174$& $7.1\pm0.1$& 7-12\\
NGC2110 & $1.65\pm0.03$& $>210$& $8.3\pm0.2$& 8-14\\
NGC5506& $1.91\pm0.03$& $720^{+130}_{-190}$& $8.0\pm0.2$& 9-2\\
NGC7213& $1.84\pm0.03$& $>140$& $8.0\pm0.2$& 10-2\\
SWIFT J2127.4+5654 & $2.08\pm0.01$& $108^{+11}_{-10}$& $7.2\pm0.2$& 11-2\\
& & & & \\
\hline
\end{tabular}
\caption{We report name of the sources, photon indices, masses and references for the whole sample. References - 1:Ballantyne et al. 2014,2:Fabian et al. 2015, 3: Matt et al. 2014, 4:Brenneman et al. 2014, 5:Balokovi{\'c} et al. 2015, 6:Marinucci et al. 2014b, 7: Parker et al. 2014, 8: Marinucci et al. 2015, 9: Matt et al. 2015, 10:Ursini et al 2015, 11: Marinucci et al. 2014a., 12:Peterson et al. 2004, 13:Ponti et al. 2012, 14:Moran et al. 2007}
\end{table*}

\section{The spectral modeling and the sample}
We hereby present the sample and the fitting techniques that led to recent high-energy cutoff measurements. Our aim is not to re-analyse the already published data but to present and discuss the values in the recent literature.

In order to infer reliable and accurate values for the cutoff energies one has to pay attention to both the primary and reflection continua. On this purpose, we show in Fig. 1 the effect that different high-energy exponential cutoffs (in particular E$\rm _c$=50, 100, 150 keV) have on the spectral curvature, in the 100--300 keV energy range. In the left-hand panel, we show three different power law (with $\Gamma=1.8$) continua, for a source with a 2--10 keV flux of F$_{2-10}=10^{-11}$ erg cm$^{-2}$ s$^{-1}$, absorbed by a column density of N$\rm _H=5\times 10^{22}$ cm$^{-2}$. 

Many bright sources in which a cutoff energy has been measured (or a lower limit has been found) show spectral features that can be ascribed to relativistic effects, arising from the innermost region of the accretion disc, in addition to cold reflection. Sources like 1H0707-495, MCG 6-30-15, Mrk 335, SWIFT J2127.4+5654 have shown intense emission from the inner side of the accretion disc when they have been recently observed by {\it NuSTAR} (Kara et al. 2015, Marinucci et al. 2014b, Parker et al. 2014, Marinucci et al. 2014a). Relativistic effects can be probed using the broad component of the iron K$\alpha$ line (observed for the first time by Tanaka et al. 1995) and the low energy extension of its red tail is an indicator of the inner radius of the accretion disc and thus of the black hole spin (Iwasawa et al. 1996, 1999, but also see Reynolds 2014 for a recent review). Both the broad (relativistic) and the narrow (emitted by cold, distant matter) components of the iron K$\alpha$ line are accompanied by Compton scattered emission peaking at $\sim 30$ keV (Matt et al. 1991; George \& Fabian 1991), in the middle of the {\it NuSTAR} energy band. Therefore, the disentanglement of these two spectral components from the nuclear continuum is fundamental to detect a curvature of the primary spectrum at high-energies.
If the nuclear continuum, illuminating the surface of the accretion disc, presents an exponential rollover at high energies, this has to be taken into account. For this purpose the majority of the works discussed in this paper use updated reflection model such as {\sc xillver} (Garcia et al., 2010, 2011, 2013) and {\sc relxill} (Garcia \& Dauser et al., 2014) that permit to take into account a high-energy curvature in both the nuclear and reprocessed spectra. We show, in the left-hand panel of Fig. 1, three different reflection spectra (using {\sc xillver} in {\sc xspec}) for a $\Gamma=1.8$ power-law illuminating continuum, with a ionization parameter $\xi=100$ erg cm s$^{-1}$ and solar abundances, for three different cutoff energies (E$\rm _c$=50, 100, 150 keV) for the same flux level used above.

\begin{figure*}
\centering
\includegraphics[width=0.6\paperwidth]{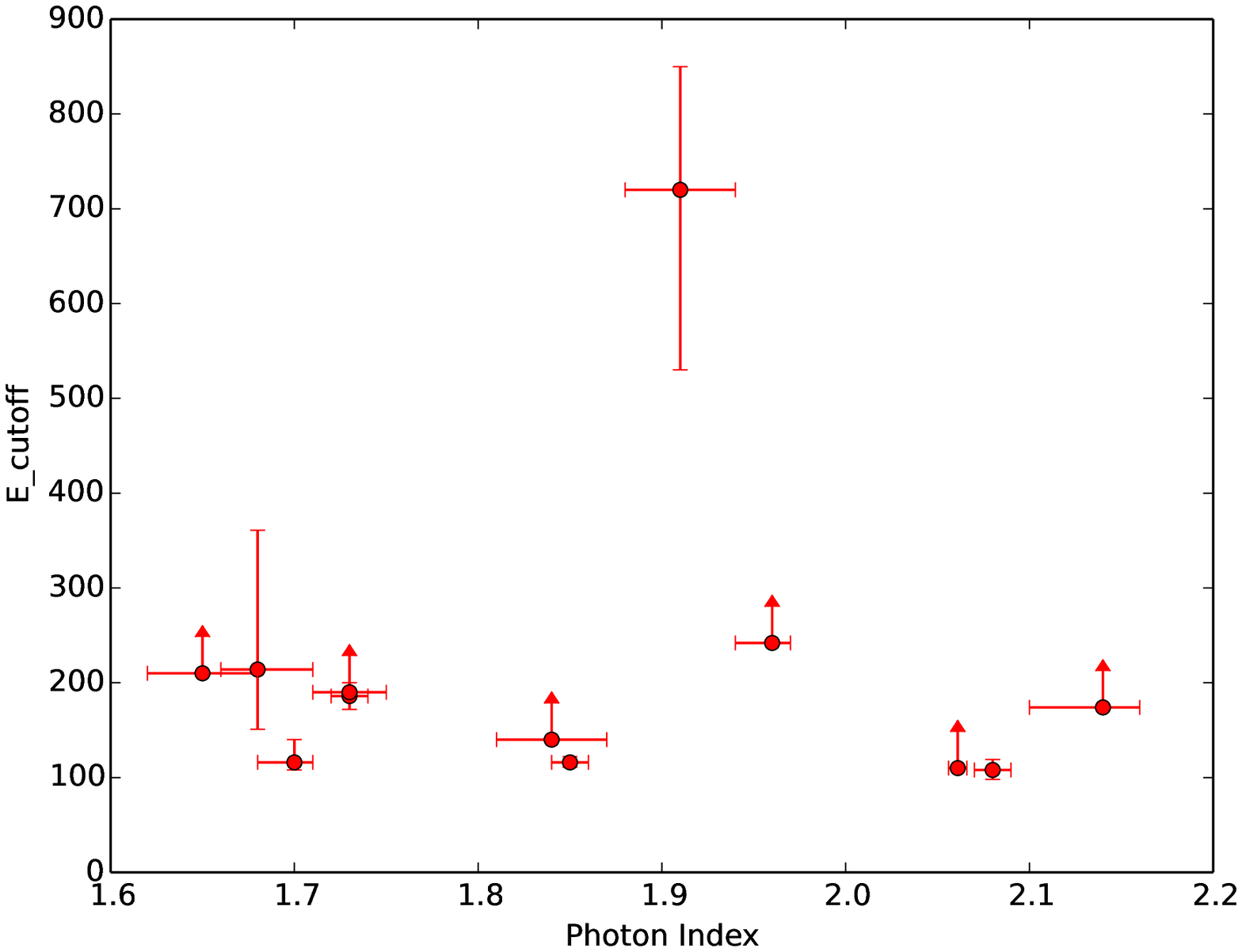}
\includegraphics[width=0.6\paperwidth]{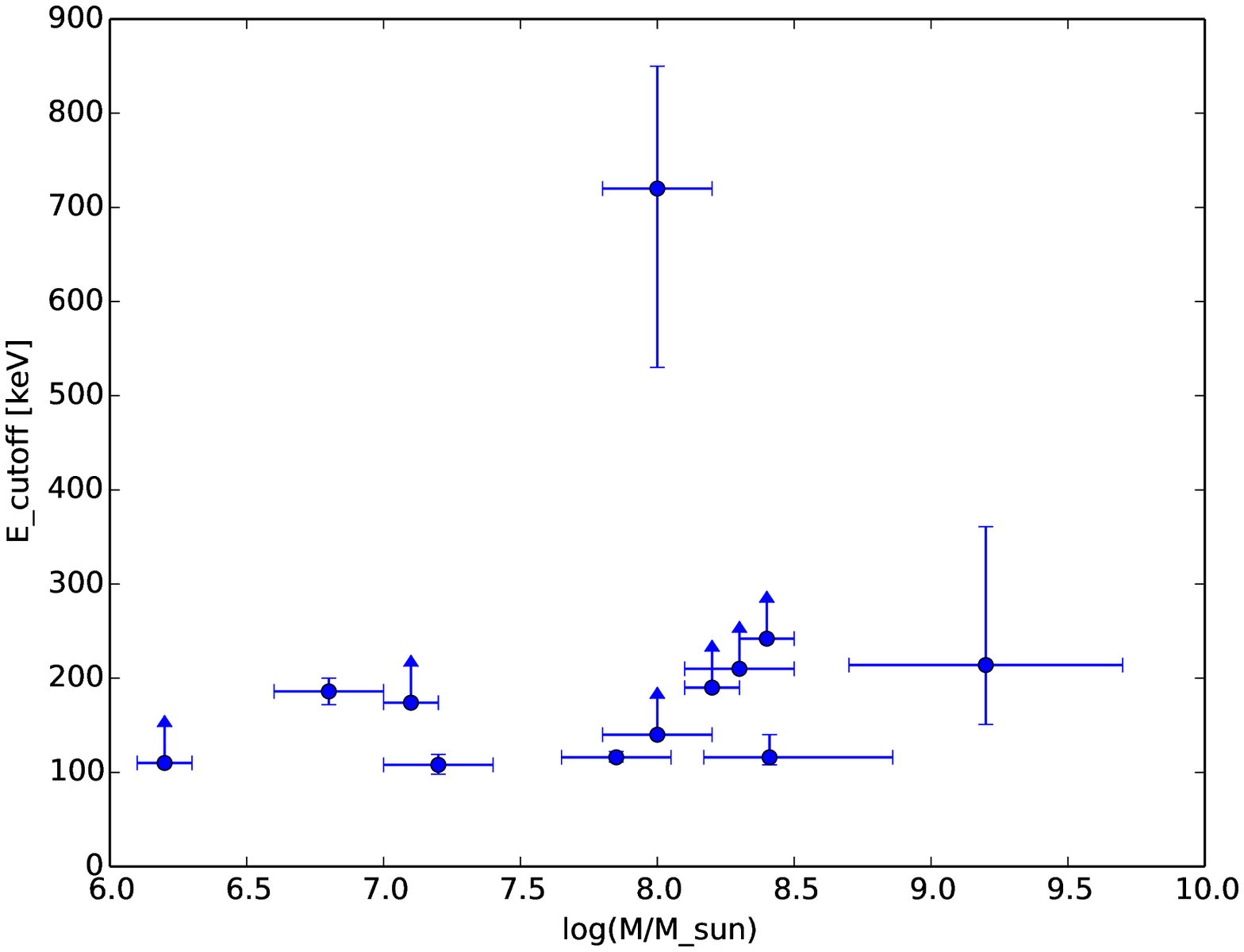}
\caption{High-energy cutoffs are plotted against the measured photon index (top panel, red data points) and estimated black hole masses (bottom panel, blue data points).  No particular trend is seen: cutoff energies are measured (or lower limits obtained) despite the steepness of the primary continuum or the black hole mass.  }
\label{plot}
\end{figure*}

In a number of observations  {\it NuSTAR} operated simultaneously with XMM-{\it Newton} and {\it Swift}. The spectral resolution around the iron line complex (5--10 keV) of these low-energy detectors has been paired with the unprecedented sensitivity above 10 keV, leading to the simultaneous monitoring of the broad iron line component and its Compton hump. In this way, it is possible to disentangle the contribution from the disc and from the 'pc-scale torus' to the total reflection fraction. For this purpose, we show in Fig. 2 the best fit model of the 2013 XMM+{\it NuSTAR} observation of the Narrow Line Seyfert 1 SWIFT J2127.4+5654 (Marinucci et al. 2014a) in which the three main spectral components can be seen: the nuclear continuum (in red), the relativistic reflection from the inner radii of the accretion disc (in purple) and the neutral reflection arising from distant material (in blue). The control upon each spectral components allowed us to break the degeneracy between the cutoff energy and several parameters, i.e. the photon index, the fraction of Compton scattered continuum (from the disc and from the 'pc-scale torus').

In the following, we choose from the literature the high-energy cutoffs measurements or lower limits in nearby AGN (z$<0.06$) that have been recently published. The sample mainly includes unobscured sources (i.e. with N$\rm _H\leq6\times 10^{22}$ cm$^{-2}$) which have been observed by {\it NuSTAR} since June 2012. Other objects for which the cutoff energy had been left fixed in the spectral analysis are not included (1H0707-495 for instance), since they need a more intensive study on this issue. We use black hole masses inferred via reverberation studies (mainly from Peterson et al. 2004) for type 1 AGN or, for the ones not present in this catalog, from Fabian et al. 2015, Ponti et al. 2012, Moran et al. 2007. We assumed a conservative 20\% uncertainty for black hole mass estimates not inferred from reverberation. The final sample is comprised of twelve objects (Table 1), with some sources in which clear evidence of both cold and relativistic reflection are present and others in which only distant neutral reflection contributes to the Compton hump at high energies. This is only a first attempt to investigate possible correlations between the cutoff energies and other physical observables such as the slope of the primary continuum and the black hole mass, a more detailed study will be presented in Tortosa et al. (in preparation).

\section{Results}
We plot our results in Fig. \ref{plot}: high-energy cutoffs are plotted against the measured photon index (top panel, red data points) and estimated black hole masses (bottom panel, blue data points). We can clearly see the improvement with respect to previous studies performed  with {\it Beppo-SAX} (Perola et al. 2002, Dadina et al. 2007) and {\it INTEGRAL} (Panessa et al. 2011, De Rosa et al. 2012, Molina et al. 2013), in which background-dominated X-ray detectors had been used. The correlation between the measured cutoff energy and the slope of the primary power law is no longer present due to unprecedented {\it NuSTAR} sensitivity at high-energies (a factor $\sim$100 gain at 20 keV with respect to {\it Suzaku} HXD-PIN: Fig. 1 in Marinucci et al. 2014): high-energy curvatures have been measured (or lower limits have been obtained) despite the steepness of the spectra. We do not find any trend also when the cutoff energy is plotted against the black hole mass.

Most of the works cited in this paper applied Comptonization models (in particular {\sc comptt}:Titarchuk 1994 , {\sc compps}: Poutanen \& Svensson 1996, {\sc nthcomp}: Zycki, Done \& Smith 1999) to reproduce the observed spectral shape of the continuum. As described above slope and high-energy cutoff of the nuclear power law are functions of the temperature of the scattering electrons in the corona $\rm kT_e$ and its optical depth $\tau$. A lower limit on the cutoff energy might be an indicator of a very high temperature electrons population and therefore of a coronal plasma in which the Compton cooling mechanism is not efficient. These models mainly use two different geometries for the corona: a spherical cloud of electrons and a semi-infinite slab above the accretion disc. The optical depth is taken vertically  for the latter geometry and radially for the former, leading to an extra 1$/cos(60)=2$ in the spherical scenario (see, for instance, discussion in Brenneman et al. 2014 for the case of IC 4329A, and references therein). 
For the case of NGC 7213, Ursini et al. 2015 recently found a high-energy cutoff of E$\rm _c>140$ keV and, assuming a spherical geometry, tested the Comptonization model {\sc compps} on the data set. The best fitting scenario is indeed a very high coronal temperature ($\rm kT_e=295^{+70}_{-250}$ keV), with an optical depth $\tau=0.2^{+0.7}_{-0.1}$.

One interesting note must be stressed for these objects in which a lower limit for the cutoff energy is found (and therefore a very high coronal temperature). Three of the twelve sources discussed here showed lower limits for E$\rm _c$ (or extremely high values)  and have very low accretion rates: NGC 5506 (E$\rm _c =720^{+130}_{-190}$ keV: Matt et al. 2015), NGC 2110 (E$\rm _c >210$ keV: Marinucci et al. 2015) and NGC 7213 (E$\rm _c>140$ keV: Ursini et al. 2015). The corresponding accretion rates (calculated as $\rm \dot{m}=L_{Bol}/L_{Edd}$) are $\rm \dot{m}=0.007-0.14$, $\rm \dot{m}=0.0025-0.037$ and $\rm \dot{m}\sim0.001$, respectively and we speculate that, in this particular objects, the low accretion rate leads to an inefficient Compton cooling mechanism. A more detailed statistical analysis of the whole sample will be discussed in Tortosa et al. (in preparation).

\section{Conclusions}
We presented and discussed the recent high-energy cutoff measurements obtained with the {\it NuSTAR} satellite, in a sample of 12 local AGN. We showed that its unprecedented operating energy range (3-79 keV) and sensitivity above 10 keV allowed to break the degeneracy between the cutoff energies and the slope of the primary continua, which had been observed with previous background-dominated hard X-rays satellites ({\it Swift-BAT}, {\it BeppoSAX}, {\it INTEGRAL}).

We emphasized that the spectral analysis has to include both the nuclear continua, from the corona itself, and the reprocessed emission from the circumnuclear environment, taking into account reflection models that include high-energy cutoffs.

We showed that no particular trend is seen, in our small sample of twelve objects: cutoff energies are measured (or lower limits obtained) despite the steepness of the primary continuum or the black hole mass. Sources with lower limits on the cutoff energy show very high coronal temperature, when Comptonization models are applied. In particular, three out of the twelve objects show a very hot corona and a low accretion rate, indicating that in these sources a possible Compton cooling inefficiency may play a role.

More observations in the next future on different objects will help us in understanding the physics of the corona--accretion disc system and its geometry.


\acknowledgements
AM and AT acknowledge financial support from Italian Space Agency under grant ASI/INAF I/037/12/0-011/13 and from the European Union Seventh Framework Programme (FP7/2007-2013) under grant agreement n.312789. This work was supported under NASA Contract No. NNG08FD60C, and made use of data from the {\it NuSTAR} mission, a project led by the California Institute of Technology, managed by the Jet Propulsion Laboratory, and funded by the National Aeronautics and Space Administration. We thank the {\it NuSTAR} Operations, Software and Calibration teams for support with the execution and analysis of these observations.  This research has made use of the {\it NuSTAR} Data Analysis Software (NuSTARDAS) jointly developed by the ASI
Science Data Center (ASDC, Italy) and the California Institute of Technology (USA).

\newpage

\appendix

\end{document}